\documentclass[%
reprint,
showpacs,
amsmath,amssymb,aps,
]{revtex4-1}

\usepackage{graphicx}
\usepackage{dcolumn}
\usepackage{bm}

\newcommand{\md}{\mathrm d}
\newcommand{\mgs}{\mu_{\rm g}^2}

\begin{document}


\title{Possible interpretation on the origin of four-fermion
interaction}%

\author{Hiroaki Kohyama}
\affiliation{Department of Physics,
National Taiwan University, Taipei 10617, Taiwan}

\date{\today}

\begin{abstract}
We present a possible interpretation on the origin of the
four-fermion interaction used in effective field theories.
Inspired by the sharp momentum peak seen in Bose-Einstein
condensate state, we incorporate the special gluon condensate
effect into the gluon propagator. We then find that, if one considers
hypothetic situation with the condensed gluon, the four-fermion
contact interaction can arise from the first principle theory
of quantum chromodynamics.
\end{abstract}


\maketitle


\section{\label{sec:intro}%
Introduction}
Quantum chromodynamics (QCD) is known to be the fundamental
theory of quarks and gluons whose ultimate goal is to describe all
the phenomena observed on hadrons, such as the proton and mesons.
Perturbative approach in QCD works well at high energy thanks
to the nature of asymptotic freedom~\cite{Gross:1973id}. However, 
it is difficult to investigate the physics at low energy where the strong
coupling becomes large. Then people often use some effective models
of QCD which contain nontrivial four-fermion
interactions~\cite{Nambu:1961tp,Gross:1974jv}.

The purpose of this letter is to discuss how this four-fermion
interaction occurs from the first principle theory of quantum field
theory. Motivated by the fact that the momentum have the sharp
peak in a Bose-Einstein condensate of gaseous matter, we incorporate
this momentum distribution into the propagator of gluons. Under the 
assumption, we find that the four-fermion contact interaction appears
as the consequence of the condensate gluon momentum.

The paper is organized as follows. We present our motivation
in Sec.~\ref{sec:motivation}. We then show the four-fermion
interaction can be derived starting from QCD in
Sec.~\ref{sec:four_fermi}. Section.~\ref{sec:model} gives a
numerical test with gluon condensate. Then some discussion
and concluding remarks are put in Secs.~\ref{sec:sde}
and~\ref{sec:conclusion}.

\section{\label{sec:motivation}%
Motivation}
\begin{figure}[h!]
\begin{center}
   \includegraphics[width=7.5cm,keepaspectratio]{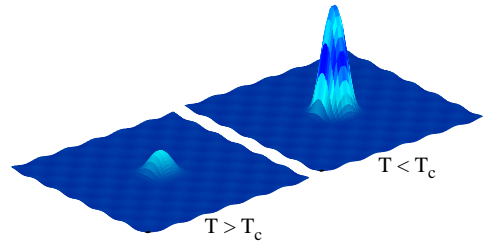}
   \caption{\label{fig:BEC}
     Schematic image of the velocity-distribution in a
     BEC state.
   }
\end{center}
\end{figure}

A Bose-Einstein condensate (BEC) is the interesting physical state
of matter which can be realized at extremely low temperature,
e.g., 170 nanokelvin for gas of rubidium atoms. The typical atomic
scale corresponds to $\AA \sim 10^{7}$K so the temperature
is indeed ultimately low. Under such the extreme circumstance, the
particles involved share their phase information, then act as if
they are one large quantum particle (this is the reason why
mean-field approximation works well for condensed matter system).
The characteristic observation on a BEC is the realization
of the momentum peak shown in Fig.~\ref{fig:BEC}
where it exhibits the image of the velocity-distribution of typical
gaseous BEC matter.

We postulate, in this letter, that the gluons in hadronic state
may have a similar tendency on the momentum distribution,
i.e., the momentum is sharply condensed around typical QCD
scale, $\mu_{\rm QCD} \sim 1{\rm fm}^{-1}(\sim 200$MeV
$\sim 10^{12}$K). Note that the room temperature ($\sim10^2$K)
is extremely low compare to the hadron scale.
We then apply this speculation by modifying the form of the
gluon propagator in QCD calculation. This is the main motivation
of the paper.

\section{\label{sec:four_fermi}%
Four-fermion interaction}
We consider the model treatment through evaluating the partition
function of QCD under the special condition mentioned in
the previous section. Thereafter we try to find the relation
between the original QCD Lagrangian and resulting effective
model, in particular, the relation between the quark-gluon
interaction and an effective four-fermion contact interaction.

\subsection{\label{subsec:QCD}%
QCD partition function}
We first review the evaluation of the partition
function  in QCD whose Lagrangian density is given by
\begin{align}
& \mathcal{L}_{\rm QCD}
  = \overline{\psi} (iD_\mu \gamma^\mu -m) \psi
                -\frac{1}{4}  F^a_{\mu\nu} F^{a\mu\nu},
\label{L_qcd}
\end{align}
where $\psi$ and $m$ are the quark field and its mass,
$F_{\mu \nu}$ is the field strength tensor and  $D_\mu$ indicates
the covariant derivative defined by
\begin{align}
 & F_{\mu\nu}^a = \partial_{\mu}A_{\nu}^a
     - \partial_{\nu}A_{\mu}^a + g f^{abc} A^b_{\mu} A^c_{\nu}, \\
 & D_{\mu}= \partial_{\mu} - i g A_{\mu}^a t^a,
\end{align}
with the gluon field $A_\mu$ and the coupling constant for the
strong interaction $g$. In this article, we follow the notations used in
the textbook by Peskin and Schroeder~\cite{Peskin:1995ev}.
It may be useful to separate the Lagrangian into the free and
the interacting parts as
\begin{align}
 & \mathcal{L}_{\rm QCD} = \mathcal{L}_{\rm q}^0 
  + \mathcal{L}_{\rm g}^0
 + \mathcal{L}_{\rm I}, \\
 & \mathcal{L}_{\rm q}^0
   = \overline{\psi} (i\partial\!\!\!/ -m) \psi, \\
 & \mathcal{L}_{\rm g}^0
   = -\frac{1}{4}(\partial_\mu A^a_\nu - \partial_\nu A^a_\mu)^2,\\
 & \mathcal{L}_{\rm I}
      = g \overline{\psi} \gamma^\mu t^a \psi A^a_\mu 
      - gf^{abc}(\partial_\mu A_\nu^a) A^{\mu b}A^{\nu c} \nonumber \\
  & \qquad -\frac{1}{4} g^2
        (f^{eab}A^{a}_\mu A^{b}_\nu)(f^{ecd}A^{\mu c}A^{\nu d}),
\end{align}
and we also use the notation $\mathcal{L}_0 =
\mathcal{L}_{\rm q}^0 + \mathcal{L}_{\rm g}^0$.
This separated form helps us to write the partition function
by the Taylor expansion
\begin{align}
  & \mathcal{Z}_{\rm QCD}
   =
    \int \! \mathcal{D}\psi \int \! \mathcal{D}\! A 
    \exp \left[
      i\int \md^4 x 
      \mathcal{L}_{\rm QCD}
       \right] \nonumber \\
   & =
      \int \! \mathcal{D}\psi \int \! \mathcal{D}\! A 
      e^{i\int \md^4 x{\mathcal L}_0} 
      \sum_{n=0}^{\infty}
      \frac{1}{n!}
      \left(  i \int \md ^4 x  \mathcal{L}_{\rm I} 
      \right)^n
\end{align}
where we will consider the terms up to the order of
$g^2$ in this article.

Before proceeding the further calculations, we define the
notation $\langle O \rangle$ for later convenience,
\begin{equation}
  \langle O \rangle
  =
      \frac{\int \mathcal{D}\psi \int \mathcal{D}\! A 
      e^{i\int \md^4 x{\mathcal L}_0} \left[ O \right]}
      {\int  \mathcal{D}\psi \int  \mathcal{D}\! A 
      e^{i\int \md^4 x{\mathcal L}_0}},  
\label{eq:expec}
\end{equation}
here Eq.~(\ref{eq:expec}) indicates the expectation value of $O$.

As is well known, quarks and gluons have never been observed
as free particles due to non-trivial effect of the confinement,
we expect the amplitudes containing the outgoing quarks and
gluons vanish, namely,
\begin{align}
  & \langle \overline{\psi} \gamma^\mu t^a \psi A^a_\mu
     \rangle
   = 0, \\
  & \langle (\partial_\mu A_\nu^a) A^{\mu b}A^{\nu c}
     \rangle
   = 0.
\end{align}
Retaining non-vanishing contributions in the partition
function, we have
\begin{align}
   &\mathcal{Z}_{\rm QCD}
     \simeq
      \int \! \mathcal{D}\psi \int \! \mathcal{D}\! A 
      e^{i\int \md^4 x{\mathcal L}_0} \nonumber \\
   & \quad \times
      \biggl[  1 + 
                 \frac{1}{2} \left( ig \int \md ^4 x  
                    \overline{\psi} \gamma^\mu t^a \psi A^a_\mu  
      \right)^2  \nonumber \\
   &   \qquad   
              + \frac{1}{2} \left( ig \int \md ^4 x  
                   f^{abc}(\partial_\mu A_\nu^a) A^{\mu b}A^{\nu c}  
      \right)^2  \nonumber \\
   & \qquad -\frac{1}{4} 
                    \left( i g^2 \int \md ^4 x 
                            (f^{eab}A^{a}_\mu A^{b}_\nu)(f^{ecd}A^{\mu c}A^{\nu d})
                    \right)
      \biggr].
\label{eq:Z_QCD}
\end{align}
This is the partition function of QCD at $g^2$ order, and we will
try to consider this quantity under special circumstance in the
following.

\subsection{\label{subsec:four_fermion}%
Quark sector}
In this subsection, we are going to study what happens if
one integrates out the gluon degree of freedom, then discuss
the possible relation between original QCD interaction and the
four-fermion interaction.

The term relating to the quark-gluon interaction
can be evaluated by
\begin{align}
   &\mathcal{Z}_{{\rm qq}g}
   = -\frac{g^2}{2} 
      \int \! \mathcal{D}\psi \int \! \mathcal{D}\! A 
      e^{i\int \md^4 x{\mathcal L}_0} \nonumber \\
   & \quad \times
      \left(
          \int \md ^4 x  
          \overline{\psi}_x \gamma^\mu t^a \psi_x A^a_{\mu x}  
          \int \md ^4 y  
          \overline{\psi}_y \gamma^\nu t^b \psi_y A^b_{\nu y} 
      \right),
\label{eq:Z_qqg}
\end{align}
where we introduced the abbreviated notations
$\psi_x = \psi(x)$ and $A^a_{\mu x} = A^a_\mu (x)$.
The usual rule for the gluon propagator reads
\begin{align}
      \bigl\langle
           A^a_{\mu} (x)  
           A^b_{\nu} (y)
      \bigr\rangle
      =
      \int \frac{\md^4 p}{(2\pi)^4}
      \frac{-ig_{\mu \nu} \delta^{ab}}{p^2}
      e^{-i p \cdot (x-y)},
\end{align}
in the Feynman gauge $\xi = 1$. Up to here, the treatment is
general; we just briefly reviewed quantum field theory
calculation. In the following, we will introduce some crude
hypothesis.

Assuming the situation that the gluon is highly condensed and its
momentum has narrow peak around $\mu_{\rm g}$ as discussed
in Sec. \ref{sec:motivation}, we perform the following replacement,
\begin{align}
      \frac{1}{p^2}
      \to
      \frac{1}{\mgs}.
\label{eq:replace}
\end{align}
We regard the quantity $\mu_{\rm g}$ as the energy scale of the
gluon. Once this brute force manipulation is allowed, we have
\begin{align}
      \bigl\langle
           A^a_\mu (x)  
           A^b_\nu (y)
      \bigr\rangle
      =
      \frac{-ig_{\mu \nu} \delta^{ab} }{\mgs}      
      \delta^{(4)} (x-y).
\end{align}
This becomes our Feynman rule for the gluon propagator
in the present model. Substituting the above rule into
Eq.~(\ref{eq:Z_qqg}), we obtain
\begin{align}
   &\mathcal{Z}_{{\rm qq}g}
     = 
     {\mathcal N}_A
      \int \! \mathcal{D}\psi
      e^{i\int \md^4 x{\mathcal L}_{\rm q}^0} \nonumber \\
     & \times
      \left[
           \frac{i g^2}{2 \mgs} 
          \int \md ^4 x  
          (\overline{\psi}_x \gamma^\mu t^a \psi_x)
          (\overline{\psi}_x \gamma_\mu t^a \psi_x)
      \right].
\label{eq:Z_qqg_2}
\end{align}
where ${\mathcal N}_A$ is the overall constant from the
functional integration on gluon.
Note that the integral for $y$ disappears due to the
delta function $\delta^{(4)}(x-y)$, then $\psi(y)$ turns
out to be $\psi(x)$ in Eq. (\ref{eq:Z_qqg_2}). This may
be regarded as the reason of contact interactions
which come from the resulting delta function.

As the final step, using the approximated relation
$e^{\epsilon} \simeq 1 + \epsilon$, we put back the
resulting term inside the exponential,
\begin{align}
   \mathcal{Z}_{{\rm q}}
    &=
     {\mathcal N}_A
      \int \! \mathcal{D}\psi
      e^{i\int \md^4 x{\mathcal L}_{\rm q}^0} \nonumber \\
     & \times \exp
      \left[
           \frac{i g^2}{2 \mgs} 
          \int \md ^4 x  
          (\overline{\psi} \gamma^\mu t^a \psi)
          (\overline{\psi} \gamma_\mu t^a \psi)
      \right].
\label{eq:Z_q}
\end{align}
Since there arises no confusion in the above expression we
drop the suffix $x$ in $\psi_x(=\psi(x))$. Then we finally arrive
at the form
\begin{align}
   &\mathcal{Z}_{{\rm q}}
     = {\mathcal N}_A
      \int \! \mathcal{D}\psi  
      e^{i\int \md^4 x{\mathcal L}_{\rm q}},
\end{align}
with 
\begin{align}
    {\mathcal L}_{\rm q}
    = \overline{\psi} (i\partial\!\!\!/ -m) \psi
       + \frac{g^2}{2 \mgs} 
        (\overline{\psi} \gamma^\mu t^a \psi)
        (\overline{\psi} \gamma_\mu t^a \psi).
\label{eq:L_q}
\end{align}
Thus we obtained the effective Lagrangian with
four-fermion contact interaction.

It is interesting that the only one replacement, although it
looks awful, leads the four-fermion contact interaction.
We think this can be a possible interpretation on the origin
of the four-fermion interaction in effective field theories.

\subsection{\label{subsec:gluon}%
Gluon sector}
We consider the gluon energy in this subsection by evaluating
the contribution of the third and fourth lines in Eq.~(\ref{eq:Z_QCD}).

By using our propagator, we see that the contribution
\begin{align}
  f^{abc} f^{def} 
  \Bigl\langle
    (\partial^x_{\mu} A_{\nu x}^a) A_x^{\mu b}A_x^{\nu c} 
    (\partial^y_{\rho} A_{\lambda y}^d) A_y^{\rho e}A_y^{\lambda f} 
  \Bigr\rangle
\end{align}
from the third line in Eq.~(\ref{eq:Z_QCD}) vanishes, due to
the property of the antisymmetric tensor $f^{abc}$.
Non-zero contribution occurs from the fourth line, in which one
sees
\begin{align}
  {\mathcal Z}_{\rm QCD}^{(4)} 
  = -{\mathcal Z}_0 \cdot 72 g^2 i 
      \int \md^4 x \phi_{\rm g}^2.
\end{align}
where ${\mathcal Z}_0 \equiv \int  \mathcal{D}\psi \int 
\mathcal{D}\! A e^{i\int \md^4 x{\mathcal L}_0}$ and $\phi_{\rm g}$
is the gluon one-loop amplitude,
\begin{align}
  \phi_{\rm g} 
  =  \int \frac{\md^4 p}{(2\pi)^4}
      \frac{-i}{\mgs},
\label{eq:g_cond}
\end{align}
with the propagator for the gluon field set by Eq.~(\ref{eq:replace}).
As obviously seen from the equation,  this one-loop contribution,
$\phi_{\rm g}$, badly  diverges, then one needs to perform the
renormalization to obtain finite physical quantity.

Putting back the resulting term into the exponential as done in the
fermion case using the trick ($1+\epsilon \to e^{\epsilon}$), we
obtain the following form
\begin{align}
  {\mathcal L}_{\rm g} 
  = 
    -\frac{1}{4}(\partial_\mu A^a_\nu - \partial_\nu A^a_\mu)^2
    -72 g^2 \phi_{\rm g}^2.
\end{align}
This is our effective Lagrangian for the gluon sector. We will
not make further analyses on this form because our focus here
is to construct the four-fermion quark model.

\section{\label{sec:model}%
Numerical test}
We have calculated the effective Lagrangian in the previous
section, and it may be now ready for performing the actual
numerical analyses.

As a simple test, we draw the phase diagram of the chiral
phase transition on temperature and chemical potential plane.
The model with the form Eq.~(\ref{eq:L_q}) is the
NJL-type model, and we just follow the prescriptions in
preceding analyses~\cite{Klevansky:1992qe, Hatsuda:1994pi}.
Applying the mean-field approximation after the Fiertz
transformation to Eq.~(\ref{eq:L_q}) in the massless 
two-flavor version, we have
\begin{align}
    {\mathcal L}
    = \overline{\psi} (i\partial\!\!\!/ -M) \psi
        -\frac{2 g^2}{9 \mgs} \phi_{\psi}^2,
\label{eq:L_NJL}
\end{align}
where $M$ is the constituent (dynamical) quark mass
$M = -2G\phi_{\psi}$ with $G=2g^2/(9\mgs)$,
and $\phi_{\psi}$ represents the chiral condensate,
\begin{align}
    \phi_{\psi}
    =
    \langle \overline{\psi} \psi \rangle
    =
    - {\rm tr}   \int \frac{\md^4 q}{(2\pi)^4}
    \frac{i}{q \!\!\!/ - M}.
\label{eq:trS}
\end{align}
In the above expression, the trace runs for the color, flavor and
spinor spaces, then the relation
$\langle \overline{\psi} \psi \rangle = \langle \bar{u}u
+ \bar{d}d \rangle$, namely, $\phi_\psi =\phi_u + \phi_d$ holds.
Here we treat $\mu_{\rm g}$ as the model parameter being some
constant, then we no longer have the renormalizability of the
original theory.

The evaluation of the effective potential at finite temperature ($T$)
and chemical potential ($\mu$) is straightforward due to the
mean-field approximation, and we have
\begin{align}
  &{\mathcal V}
  = \frac{2 g^2}{9\mgs} \phi_{\psi}^2
   - 12 \int \frac{\md^3 q}{(2\pi)^3}
    \left[ E + T \sum_\pm \ln \left( 1 + e^{- \beta E^\pm} \right)
    \right],
\end{align}
with $E^\pm = E\pm\mu$, $E=\sqrt{q^2 + M^2}$ and
$\beta = 1/T$~\cite{Huang:2004ik}.
Once we have the effective potential, the expectation value
of the order parameter, $\phi_\psi$, can be determined by the
gap equation, 
$ \partial \mathcal V/ (\partial \phi_\psi) =0$,
corresponding to the stational condition of the effective potential.

The remaining preparation for the numerical analysis is the
parameter fitting. As usual, we introduce the three-momentum
cutoff, $\Lambda$, to obtain the finite contribution from the loop
integral. Thereafter the present model contains three parameters, the
strong coupling $g$, the three momentum cutoff $\Lambda$
and the gluon energy scale $\mu_{\rm g}$. Here we chose,
$\Lambda=623$MeV and $g = 1.27$, then test various values
of $\mu_{\rm g}$. The above parameters are chosen so that
the model reproduces the numbers, $M=335$MeV and
$f_\pi=92$MeV for $\mu_{\rm g}=250$MeV.

\begin{figure}[h!]
\begin{center}
   \vspace{-0.2cm}
   \includegraphics[width=7.5cm,keepaspectratio]{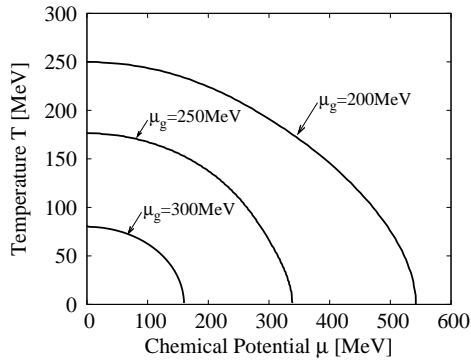}
   \vspace{-0.2cm}
   \caption{\label{fig:pd}
     $\mu_{\rm g}$ dependence on the phase diagram.
   }
   \vspace{-0.2cm}
\end{center}
\end{figure}

Figure~\ref{fig:pd} displays the numerical results of the phase
diagram on the chiral phase transition.
One sees that the region of the broken phase shrinks with
increasing $\mu_{\rm g}$. This can easily be understood, because
the effective coupling strength, $G(\mgs) = 2g^2/(9\mgs)$,
becomes weak when $\mgs$ is larger. Observing the gluon scale
dependence on the model, we think that $\mu_{\rm g}$ plays a
similar role with the renormalization scale in quantum field theory.

\section{\label{sec:sde}%
Other relation}
It may also be worth mentioning that the relation between the
Schwinger-Dyson equation (SDE)~\cite{Dyson:1949ha,
Schwinger:1951ex} and the NJL can be seen by
a similar way. Below shows the SDE for the dynamical mass,
$M(p)$,
\begin{align}
    M(p)
    =-i C_2 g^2
     {\rm tr}   \int \frac{\md^4 q}{(2\pi)^4}
    \frac{-i}{k^2}
    \frac{i}{q \!\!\!/ - M(q)}
\label{eq:SDE}
\end{align}
with $C_2 = (N_{\rm c}^2-1)/(2N_{\rm c})$ and $k=p-q$,
where we chose the Feynman gauge and set the field strength
renormalization factor to be unity in the original equation for
explanation simplicity. It should be noted that the trace  does
not include the flavor space in Eq.~(\ref{eq:SDE}) contrary to
Eq.~(\ref{eq:trS}). Performing the replacement, $1/k^2 = 
-1/\mgs$, we see
\begin{align}
    M(p)
    =C_2 \frac{g^2}{\mgs}
     {\rm tr}   \int \frac{\md^4 q}{(2\pi)^4}
    \frac{i}{q \!\!\!/ - M(q)}.
\label{eq:SDE2}
\end{align}
Further, if one drops the momentum dependence on $M(p) \to M$
and recalls the relation $M_u = -4G \phi_u$ for up quark,  
\begin{equation}
    \phi_u
    =- \frac{9C_2}{8}
     {\rm tr}   \int \frac{\md^4 q}{(2\pi)^4}
    \frac{i}{q \!\!\!/ - M_u}.
\label{eq:SDE3}
\end{equation}
Thus we just get the same form with the NJL gap equation.
Note that if the models are numerically close under the assumption
of constant $k^2$ and $M(p)$, the relation for the coupling strength
is expected to be $ g_{\rm SDE}^2 / g_{\rm NJL}^2 \simeq 2/3$.
Although the direct comparison is difficult, practical analyses show
both couplings may have similar values~\cite{Roberts:1994dr}.

\section{\label{sec:conclusion}%
Concluding remarks}
We find that the frequently studied four-fermion
interaction in effective models can be derived by QCD, if we
employ the brute forth hypothesis shown in Eq.~(\ref{eq:replace}).
The performed manipulation is based on the speculation that
the gluon momentum may have narrow peak as seen in 
a BEC state in condensed matter physics. We believe that,
since the simple replacement can produce the four point
interaction, the treatment employed here may have
some physical importance.

Since the NJL model, an effective model of QCD, is introduced
by using the analogy from the Bardeen Cooper Schrieffer (BCS)
theory~\cite{Bardeen:1957kj}, an effective theory of quantum 
electrodynamics (QED). Therefore, we believe that the relation 
between QED and the BCS theory can be read in a similar 
manner presented in this paper.

\begin{acknowledgments}
The author thanks to T. Inagaki for discussions.
The author is supported by Ministry of Science and Technology
(Taiwan, ROC), through Grant No. MOST 103-2811-M-002-087.
\end{acknowledgments}





\end{document}